# A Quality Model for Actionable Analytics in Rapid Software Development

Silverio Martínez-Fernández, Andreas Jedlitschka, Liliana Guzmán, Anna Maria Vollmer
*Fraunhofer IESE*
Kaiserslautern, Germany
{Silverio.Martinez, Andreas.Jedlitschka, Liliana.Guzman, Anna-Maria.Vollmer}@iese.fraunhofer.de

*Abstract*— Background: Accessing relevant data on the product, process, and usage perspectives of software as well as integrating and analyzing such data is crucial for getting reliable and timely actionable insights aimed at continuously managing software quality in Rapid Software Development (RSD). In this context, several software analytics tools have been developed in recent years. However, there is a lack of explainable software analytics that software practitioners trust. Aims: We aimed at creating a quality model (called Q-Rapids quality model) for actionable analytics in RSD, implementing it, and evaluating its understandability and relevance. Method: We performed workshops at four companies in order to determine relevant metrics as well as product and process factors. We also elicited how these metrics and factors are used and interpreted by practitioners when making decisions in RSD. We specified the Q-Rapids quality model by comparing and integrating the results of the four workshops. Then we implemented the Q-Rapids tool to support the usage of the Q-Rapids quality model as well as the gathering, integration, and analysis of the required data. Afterwards we installed the Q-Rapids tool in the four companies and performed semi-structured interviews with eight product owners to evaluate the understandability and relevance of the Q-Rapids quality model. Results: The participants of the evaluation perceived the metrics as well as the product and process factors of the Q-Rapids quality model as understandable. Also, they considered the Q-Rapids quality model relevant for identifying product and process deficiencies (e.g., blocking code situations). Conclusions: By means of heterogeneous data sources, the Q-Rapids quality model enables detecting problems that take more time to find manually and adds transparency among the perspectives of system, process, and usage.

## I. Introduction

A recent report by Capgemini states that the average spending on quality management and testing in IT companies grew from 18% in 2012 to 35% in 2015, that and this proportion of the budget is estimated to increase to 40% by 2018 [1]. Companies want to avoid discovering a software quality problem when it is already too late or too expensive to fix it. Quality management is even more crucial in each cycle of Rapid Software Development (RSD), as it allows making preventive strategic decisions (e.g., prioritization of a product backlog). RSD refers to the organizational capability to develop, release, and learn from software in rapid cycles [2].

Software analytics is a recent approach for improving software quality, development productivity, and user experience [3]. It utilizes data-driven approaches to obtain insightful and actionable information to help software practitioners with their data-related tasks [4]. As of March 2018, a significant increase in the popularity of software analytics has been noted [5]. Examples of commercial tools include Microsoft Azure Application Insights, Codacy, Seerene, and Revulytics. Regarding academic tools, examples are SQUALE [6], QuASE [7], CodeFeedr [8], and the Q-Rapids tool [9] (novel in generating data-driven quality requirements from both runtime and design-time data [10]). However, as argued by Dam et al.: "One of the key reasons is that software practitioners are reluctant to trust predictions produced by the analytics machinery without understanding the rationale for those predictions" [11].

Our software analytics research is part of the Q-Rapids European H2020 research project about managing quality during rapid software development. The goal of our work is to provide explainable software analytics so that decision makers will trust our information when it comes to identifying quality deficiencies, planning countermeasures, and determining quality requirements in the context of RSD. For this reason, we have created a quality model – hereafter referred to as the Q-Rapids quality model – based on the quality needs elicited during workshops with practitioners in four companies. Asking practitioners is a more time-consuming process than only collecting and analyzing data, as most of the existing software analytics tools do. In return, however, it improves the understandability of the quality model and fosters the resolution of essential problems as understood by practitioners.

Other features of the Q-Rapids quality model are the integration of heterogeneous data sources, the aggregation and interpretation of raw data, and the selection of relevant and business-oriented metrics. First, the quality model integrates and analyzes heterogeneous data sources during development and at runtime. This makes it possible to increase the stakeholders' awareness of the holistic quality of software, process, and usage. For instance, developers usually have a limited view of software quality, lacking updated information from runtime sources (i.e., system behavior and end user feedback). Not measuring software quality in a holistic manner leads to software quality issues being communicated subjectively to other stakeholders. Second, the quality model aggregates the collected raw data in order to provide an assessed quality overview that focuses only on general quality aspects. When a quality aspect is at risk, the model enables the

This work has been supported by the Q-Rapids H2020 European project (no. 732253), and the ERCIM Fellowship programme.

generation of quality alerts (i.e., the interpretation that a quality aspect does not have the desired value). In the case of a quality alert, the stakeholders can inspect the details of the raw data and look for actionable analytics. Third, by focusing on organization-specific indicators, only relevant metrics are highlighted so that the stakeholders know where to look for software quality issues objectively in a goal-oriented way (i.e., not wasting time by looking at irrelevant aspects).

The Q-Rapids quality model is deployed in the four industrial use cases of the Q-Rapids project. Its implementation and deployment provide the whole quality model in a single place, instead of spreading it across a variety of tools. Additionally, we performed an evaluation with eight practitioners based on real data from four use cases.

This paper is structured as follows. Section II presents a brief background on quality modeling. Section III introduces the research question and the research methodology applied. Section IV defines the Q-Rapids quality model and how it is executed and used. Section V explains how the quality model has been implemented in the Q-Rapids tool. Section VI reports the evaluation results from using the quality model in the four use cases. Finally, Section VII reports the conclusions and presents an outlook on future work.

## II. BACKGROUND

There exists a multitude of software-related quality models that are used in practice, as well as many classification schemes [12]–[14]. One example is the ISO/IEC 25010 standard [15], which determines which quality aspects to take into account when evaluating the properties of a software product. A more recent example is the Quamoco quality model [16], which integrates abstract quality aspects and concrete quality measurements. To do so, abstract *quality aspects* are broken down into *product factors* (attributes of parts of the product that are concrete enough to be measured), *assessed metrics* (concrete descriptions of how specific product factors should be quantified and interpreted for a specific context), and *raw data* (i.e., the data as it comes from the different data sources, without any modifications); see Fig. 1 for an example. Nowadays, operationalized quality models offering actionable analytics for multiple purposes (system, process, and usage) are still a challenge.

To capture essential problems and provide a measurement program, GQM offers an approach for goal-oriented measurement [17]. Starting from the goals, questions are derived. By answering these questions, respective metrics are defined for quantifying the goals. Thus, GQM provides a way to define metrics and interpret them. This allows demonstrating how to make quality aspects measurable and where to get the data. In addition, GQM+Strategies™ aligns an organization's goals and strategies across different units through measurement [18]. Besides a clear understanding of what the goals of the organization are, the use of GQM+Strategies™ facilitates communication between different units by creating a common understanding. It helps to show the developers their contribution to the higher-level key performance indicators. In our work, the use of GQM+Strategies™ allowed identifying essential business problems and relating them to the quality of product, process, and usage.

## III. RESEARCH METHODOLOGY

### A. Research Questions

We aimed at creating the Q-Rapids quality model for actionable analytics in the context of RSD. In view of this, we defined the following research question: *Which product and process factors are relevant for providing actionable analytics in RSD?*

### B. Context

The creation and evolution of the Q-Rapids quality model is driven by four use cases defined in collaboration with Nokia, Bittium, Softeam, and ITTI [2]. The use cases were selected to show the generalizability of Q-Rapids. A use case here refers to a software system developed in an RSD process in which the Q-Rapids quality model is being used to provide product owners with actionable analytics. The use cases cover the quality management of single and multiple product lines for several application domains such as telecommunication, security, military, transport, health, and public administration.

### C. Methodology

*1) Specification of Use Cases:* To specify the use cases, we asked the representative of each company to present the company setting (i.e., size, products, and type of projects) as well as the project setting (i.e., application domain, project goals, project organization, development methodology, functional requirements, quality requirements, and current quality issues). We also asked each representative to explain their expectations on actionable analytics. Each presentation was prepared following a predefined template designed by researchers of Fraunhofer IESE. This template was intended to gather the required information at a similar level of detail across the four companies. The presentations lasted an average of 60 minutes (Min = 45, Max = 65). At least two researchers documented the gathered information.

We also carried out semi-structured interviews with team members of each selected project. In total, we conducted ten interviews. The goal of the interviews was to resolve open issues and get detailed information regarding the project's products, processes, methodologies, and data sources. After documenting each use case, each representative revised the corresponding specification.

*2) Quality Workshops:* We performed one quality workshop per company aimed at eliciting: (i) the needs of managers and product owners regarding actionable analytics; (ii) relevant product and process factors as well as their metrics; and (iii) how product and process factors are used and interpreted to make decisions. The results were documented as a company-specific quality model.

Each workshop lasted eight hours and included up to ten team members of the corresponding project. Two researchers of Fraunhofer IESE moderated each workshop based on a standardized guideline combining the approaches presented in

Section II. First, we briefly introduced quality modeling and explained GQM+Strategies™ and ISO/IEC 25010. Second, we asked each participant to work individually and specify goals related to product and process factors using the GQM goal template. Then each participant explained the specified product and process factors and (if appropriate) mapped each one to ISO/IEC 25010. Third, we asked the participants to work in small groups to derive measurement goals (including metrics) from a subset of the specified product and process factors. The results were documented and discussed with all participants using the GQM abstraction sheet. Fourth, we asked each group to specify operational quality gates for each GQM abstraction sheet. An operational quality gate includes: (i) the questions to be answered in the corresponding GQM goal template; (ii) an example of the visualization of the metrics required for providing an answer as well as its interpretation; and, (iii) the necessary data sources. The group results were discussed with all participants. Finally, we summarized the workshop results.

*3) Consolidation of the Quality Model:* Based on the workshops results, we specified practitioner-relevant user stories. Moreover, we created the Q-Rapids quality model by comparing, relating, and integrating the company-specific quality models. Thus, we identified commonalities and variabilities regarding relevant product and process factors as well as metrics among the four companies. We also checked these factors and metrics regarding their feasibility for being measured automatically. Then we presented the Q-Rapids quality model in two face-to-face meetings to a subset of the participants who had attended the quality workshops. These meetings were intended to get further feedback and improve the Q-Rapids quality model. Although more metrics had been identified in the workshops, the Q-Rapids quality model includes an initial list that was applied in the four companies.

We elicited and implemented the Q-Rapids quality model between December 2016 and December 2017. In total, 20 practitioners working in RSD attended the quality workshops and were involved in the review of the company-specific quality models, and eight participants attended the face-to-face workshops to provide feedback on the Q-Rapids quality model. The Q-Rapids tool includes the implementation of the Q-Rapids quality model as well as the gathering, integration, and analysis of the required data.

IV. Q-RAPIDS QUALITY MODEL

In this section, we will describe the elements of the Q-Rapids quality model and explain how to use it for quality assessment and actionable analytics.

*A. Elements of the Q-Rapids Quality Model*

Table I shows all the elements of the Q-Rapids quality model: quality aspects, product and process factors, assessed metrics, and raw data. Next, we will further explain several product and process factors for the *maintainability*, *reliability*, *functional suitability*, and *productivity* quality aspects. The first three quality aspects are from ISO 25010 and refer to the quality of the software system. The fourth quality aspect refers to the productivity of the software development process.

*1) Code quality (product factor):* Developers, code guardians, and integrators want to gather data about the impact of code changes on code quality, so that they can manage maintainability resources. Metrics for code quality, such as complexity, comment density, and duplication density, come from static code analysis (e.g., SonarQube). *Code quality* is actionable when the product owner decides to invest a cycle into maintainability and understandability.

*2) Blocking code (product factor):* Developers, code guardians, and integrators want to gather data about code changes, so that they can identify new quality issues and blocking code. Metrics about the fulfilment/violation of quality rules and technical debt indices come from static code analysis (e.g., SonarQube). To identify risky files, these metrics should be combined with commit information from repositories (e.g., SVN, git, GitLab), such as number of lines of code modified, and how many times a file or a set of files has been jointly modified. Action points for *blocking code* include resolving blocker quality rule violations or refactoring highly changed files (e.g., God objects or configuration files).

*3) Testing status (product factor):* Test managers, quality assessors, and integrators want to gather data about the quality and stability level of testing, so that tests meaningful on the one hand and not skipped on the other hand. Example metrics for testing and integration are test coverage from static code analysis (e.g., SonarQube) and the results and the duration of tests performed by continuous integration tools (e.g., Jenkins). It is also important to differentiate between defects discovered during development validation and at runtime (e.g., from end users' complaints). Action points for *testing status* include improving tests that do not detect critical bugs during development, or improving the performance of the test pipeline.

*4) Software stability (product factor):* Product directors and quality managers want to gather data about the most critical issues at runtime, so that they can maintain efficient service capability/quality/prioritization. Metrics for crashes at runtime (e.g., type of error, mean time between failures) can be gathered from logs and network monitoring tools (e.g., Kibana, Zabbios, Nagios). It is important to indicate in the issue tracking tools which bugs are discovered at runtime. This way, statistics over time and across users can be shown, and testing effectiveness can be measured. Actionable analytics for *software stability* include the urgent generation of alerts when a fault has occurred.

*5) Software usage (product factor):* Product directors and product owners want to gather data about the product usage (e.g., total time spent on functionalities, and functionalities used most/least), so that it makes completely clear how heavily each feature is used by customers and in which order the features should be prioritized for inclusion. Metrics for the statistics of product usage, such as the number of times a feature is used, may come from log file analysis or from a customized plugin to be embedded in the product. An example of an action point for *software usage* is the removal of features that are not used in the software product.

*6) Issues' velocity (process factor):* Product owners and project managers want to gather data about content delivered at feature build compared to planned content on exit, so that they can see the planning capability and accuracy of the team. Relevant metrics for the accuracy of planning tasks are the estimated time and the actually invested time for a task from issue tracking tools (e.g., Redmine, GitLab, JIRA, Mantis). From the same data sources, metrics for the productivity of closing tickets for tasks and issues such as starting and ending date can be gathered. It is important to combine this information with commit information to aggregate data about the effort invested in code changes. For this, it is crucial that developers indicate the id of the task or issue in the commit description (e.g., git). Actionable analytics for *issues' velocity* require updates for the process, such as learning from inaccurate planning or estimation, in order to better plan the next cycles, or specifying how to use the issue tracking system.

## B. Execution of the Quality Model Assessment

Decision makers want to easily aggregate raw data from heterogeneous data sources into product and process factors as well as quality aspects. For instance, as can be seen in Table I, *reliability* is computed based on data from continuous integration systems, tests, issue tracking systems, logs, and network monitoring tools.

The assessment of the Q-Rapids quality model follows the Quamoco bottom-up approach shown in Fig. 1. Quality aspects are calculated based on product and process factors. Both product and process factors are calculated based on the assessed metrics. These are calculated from raw data, which may come from heterogeneous data sources.

The example presented in Fig. 1 illustrates how to compute the quality aspect *maintainability* according to its definition in Table I. This is done in the following six steps.

TABLE I. Q-RAPIDS QUALITY MODEL

| QA[a] | Factor[c] | Assessed Metric[c] | Description | Raw Data | Data Source |
|---|---|---|---|---|---|
| Maintainability | Code Quality | Non-complex files[b] | Files below the threshold of cyclomatic complexity (10 by default) | *Cyclomatic complexity* per *function* of each file, *total number of files* | SonarQube |
| | | Commented files[b] | Files whose comment density is outside the defined thresholds (by default 10%-30%) | *Density of comment lines* and *lines of code* per each file | SonarQube |
| | | Absence of duplications[b] | Files below the threshold of duplicated lines percentage | *Duplicated lines* and *lines of code* per file | SonarQube |
| | Blocking Code | Fulfillment of critical/blocker quality rules[b] | Files without critical or blocker quality rule violations | Number of *quality rule violations* per file and their *severity* (blocker, critical, major, minor, info) and *type* (code smell, bug, vulnerability) | SonarQube, Coverity, CodeSonar |
| | | Highly changed files | Unstable files that have been highly changed in the last commits | For each *commit*: *files changed, lines of code added/modified/deleted, author,* and *revision* | SVN, git, Gerrit |
| Reliability | Testing Status | Passed tests[b] | Unit test success density | Number of *unit test errors, failures, skipped,* and *total* | Jenkins, GitLab |
| | | Fast test builds[b] | Test builds below the duration threshold | *Duration of unit test execution, tests* conforming to a *pipeline* | Jenkins, GitLab |
| | | Test coverage | Tests with appropriate coverage | *Condition coverage* and *line coverage* per *unit test* | Jenkins (JaCoCo plugin), SonarQube |
| | Software Stability | Non-bug density[b] | Ratio of open issues of the type bug with respect to the total number of issues within a customized timeframe | Total number of *issues* (a.k.a. tasks) per *status* (e.g., open, done), *type* (e.g., bug, maintenance, feature), and *timeframe* (e.g., current/last month or current/last sprint) | Jira, GitLab, Redmine, Mantis |
| | | Errors at runtime | Occurrence of critical errors at runtime at the end user site | All tracks of logs, including *type of error* (fatal, error, warning, info, debug, trace), *file* and *line* where it occurred, and *error message* | Logs |
| | | Availability uptime | Percentage of time that the product is accessible | Timestamp at which the system is not available and derived metrics, e.g., *availability uptime, mean time between failures* | Zabbix, Nagios |
| Functional Suitability | Software Usage | Feature usage | Appropriateness of the features included in the software product regarding their usage | For each *functionality (or feature)*: number of *times used, average usage time, customer feedback on the feature* (if available) | Logs, monitoring plugin |
| Productivity | Issues' Velocity | Resolved issues assigned to a date | Resolved issues assigned to a date (simple date, iteration, or release) | For each *issue* (a.k.a. task): *time created, status, time updated, iteration(s), release(s).* | Jira, GitLab, Redmine, Mantis |
| | | Issues completely specified[b] | Density of incomplete issues within a timeframe | *Fields of each issue* (e.g., description, due date, assignee, estimated time) | Jira, GitLab, Redmine, Mantis |

[a.] QA (Quality Aspect).   [b.] The assessed metrics marked with 'b' were implemented in the quality model and evaluated in January 2018 (see Section VI). The other identified assessed metrics from the workshops have not been implemented yet.   [c.] Each assessed metric can be classified into more than one product or process factor. In the same way, each product and process factor can be classified into more than one QA.

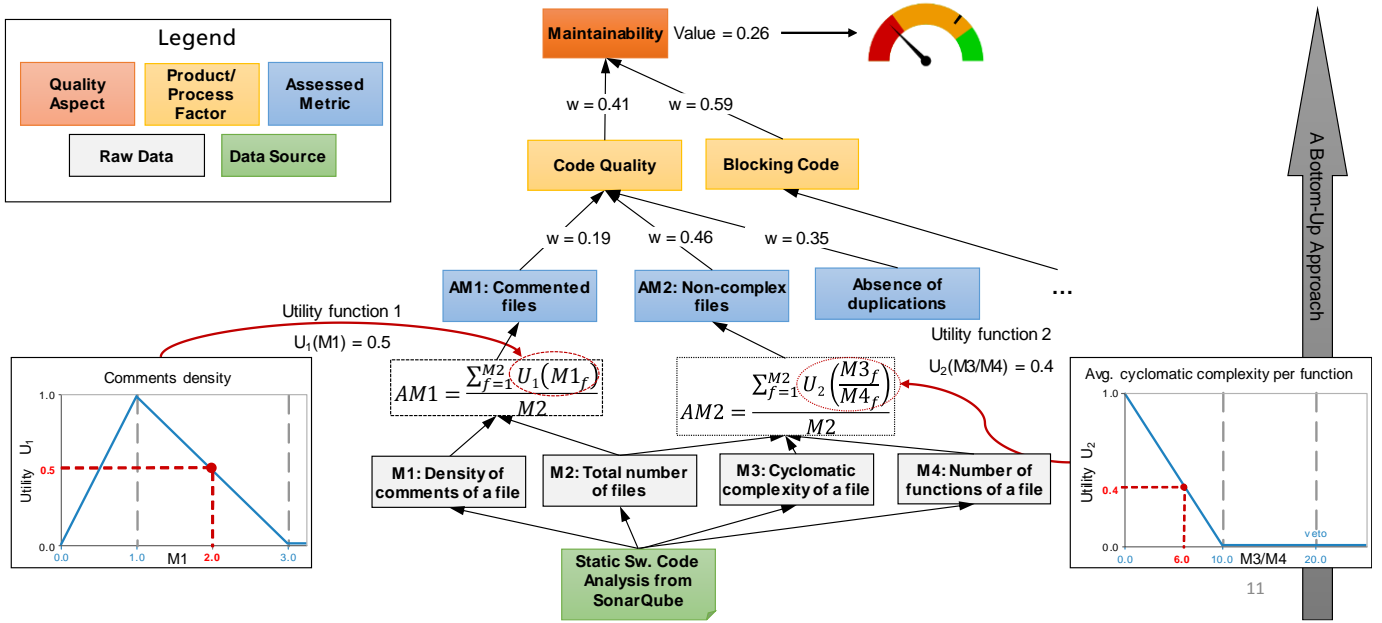

Fig. 1. Example of utility functions for assessing the "maintainability" quality aspect (adapted from Quamoco).

First, we gather information from the static code analysis executed by *SonarQube* data source. Second, we identify the basic metrics contained in the *raw data*, namely: *M1, M2, M3, M4*. Third, we need to interpret the raw data. The interpretation of raw data is performed with a *utility function*, which interprets the raw data value by using the preferences and judgments of experts and/or learned data (e.g., machine learning). Therefore, utility functions model the preferences of decision makers with respect to the data. The output of the interpretation of a basic metric is a value between 0 and 1, with 0 being the worst value and 1 the best value regarding quality. In the example in Fig. 1, we can see the utility functions $U_1$ and $U_2$ for comment density (*M1*) and average cyclomatic complexity per function (*M3/M4*) as basic metrics of *raw data*. For instance, the ideal raw value of cyclomatic complexity is 0, which is mapped to 1 (best). When the cyclomatic complexity has a value of 6, it is mapped to the interpreted or assessed utility value 0.4. Another example: when the cyclomatic complexity is equal to or greater than 10, the assessed value is 0. Fourth, the assessed metrics are computed using the raw data interpretation from the previous step. In the example, we can see that the assessed metric *non-complex files* is calculated after analyzing the *cyclomatic complexity* of all the functions of the source code by applying the utility function $U_2$ in all files (*f* sub-index). Fifth, the assessed metrics are aggregated into product or process factors (such as *code quality*) depending on their weight. The weight is determined either by experts and/or learned data. The weight qualifies the relative importance of the assessed metric for the product and process factor. Sixth, product and process factors are aggregated into quality aspects (such as *maintainability*) in the same way as in the previous step.

*C. Using the Quality Model: Quality Alerts*

As shown in Fig. 1, the three most abstract levels (quality aspects, product and process factors, and assessed metrics) work as "traffic lights", with a normalized value between 0 and 1 and customized thresholds. Hence, the users of the Q-Rapids quality model can customize at which point quality alerts should be raised. Below, we give an example of how to use our quality model with respect to the *maintainability* quality aspect:

A company has to improve *maintainability* for one of their high-quality products. In the quality model, the quality aspect *maintainability* is composed of two product factors: *code quality* and *blocking code*. In this company, Bob, a quality manager, decides to use the Q-Rapids tool to manage *maintainability* problems. He installs the tool and the quality model does not raise any alert. However, at the beginning of the next cycle, Bob receives an alert because the *maintainability* bell is ringing. He sees in the tool that the *maintainability* traffic light has moved from green to orange. He calls for a meeting with Jane, a senior developer. Together, they go deeper into the quality model to analyze the situation in depth. Although *code quality* is green, *blocking code* appears as red. They further explore the assessed metrics and raw data of *highly changed files* and *fulfillment of critical/blocker quality rules*. For evaluation purposes (see Section VI), we trained the users to only use the three most abstract levels. However, if trained to visualize the collected raw data, they could identify that in the last cycle, the classes of a directory were changed many times by a single developer. Moreover, these classes contained five violations of blocker quality rules about code smells. Raw data visualization offers actionable analytics to refactor the classes of the problematic directory, clearly indicating which classes have been heavily modified and an explanation of the violated quality rules. They could take action by adding a new issue to the backlog so that the author can solve these problems and not accumulate technical debt. As an example, see Fig. 2, where the violations of blocker quality rules are highlighted at the bottom right. In Section VI.E, we will present the improved training so that the quality model will also offer actionable analytics.

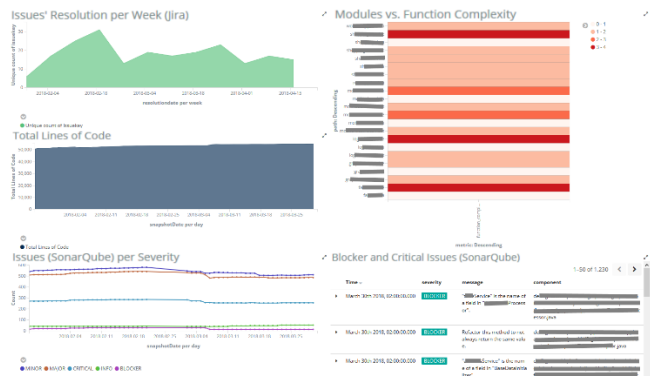

Fig. 2. Example of raw data visualization and actions for *maintainability*.

## V. IMPLEMENTATION OF THE QUALITY MODEL

This section explains how the Q-Rapids quality model has been implemented within the Q-Rapids tool [9]. Fig. 3 shows a high-level architecture view depicting the modules of the Q-Rapids tool and the data flow. It employs the idea of the Lambda architecture approach used for Big Data solutions [19]. We report below the red part of Fig. 3, composing the four layers related to the Q-Rapids quality model. For further details, the reader is referred to [9].

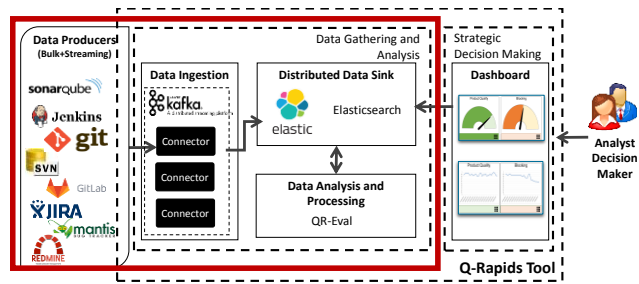

Fig. 3. Adjusted Lambda Architecture for the Q-Rapids quality model.

First, the data producers layer consists of external heterogeneous data sources with information about software quality. Currently, the Q-Rapids tool supports data gathering from static code analysis, tests executed during development, code repositories, and issue tracking tools.

Second, the data ingestion layer consists of several Apache Kafka connectors to gather data from data producers. These connectors query the API of data producers to ingest the data into Kafka. Apache Kafka is a Big Data technology serving as the primary ingestion layer and messaging platform, and offering scalability via cluster capabilities.

Third, the distributed data sink layer is used for data storage, indexing, and analysis purposes. Both the raw data (i.e., data collected from different RSD cycles) and the quality model assessment (i.e., the aggregations) are stored in a search engine called Elasticsearch. This allows defining four types of indices: three for the most abstract elements of the quality model (quality aspects, product and process factors, and metrics), and the fourth for the raw data. Like Apache Kafka, Elasticsearch offers scalability via cluster capabilities, which is required when storing huge amounts of data. Besides, we have selected the Elastic stack due to its capability to quickly perform aggregations, which becomes fundamental for the different levels of the quality model.

Fourth, the data analysis and processing layer performs the quality model assessment (see Fig. 1). The execution of the quality model assessment is performed by querying the distributed data sink and applying the utility functions in properties files to interpret the raw data. This is highly customizable for the needs of each use case. For instance, the Q-Rapids tool users can set up the quality model (utility functions, weight of product factors, and so on), and the frequency in which the quality model assessment is executed (e.g., daily, hourly). When the upper levels do not fulfill the thresholds, quality alerts are raised. Quality alerts offer actionable analytics, including raw data visualization, to help decision makers (see Fig. 2).

## VI. EVALUATION

### A. Evaluation Goals and Methodology

We aimed at characterizing the understandability and relevance of the Q-Rapids quality model from the perspective of product owners and at identifying need for improvement. We defined three evaluation questions:

Q1. Understandability of the quality model – To what extent is the Q-Rapids quality model understandable for product owners?

Q2. Relevance of the quality model – To what extent do product owners believe the Q-Rapids quality model is relevant?

Q3. Need for improvement – What needs to be improved to increase the understandability and relevance of the Q-Rapids quality model?

We had two main constraints regarding the design of the evaluation of the Q-Rapids quality model: (i) The context was predefined, namely the four use cases selected by the companies working in the Q-Rapids project; and (ii) the availability of practitioners for participation in the first evaluation was low (up to two hours per person). Thus, we performed individual semi-structured interviews with eight participants: First, we explained the study goals and procedure. Second, we trained the participant in the Q-Rapids quality model by explaining its implemented elements (see Section IV) and in the Q-Rapids tool functionalities. Third, the participant used the Q-Rapids tool to explore the Q-Rapids quality model and the underlying project data. We encouraged the participant to think aloud and mention positive aspects as well as suggestions for improvement of the Q-Rapids quality model. Fourth, we collected further feedback regarding the understandability and relevance of the Q-Rapids quality model by using a questionnaire.

We operationalized the understandability and relevance of the Q-Rapids quality model based on the definitions and questions introduced in [20] and [21], respectively. Each Likert scale included up to four statements to be rated using a response scale from 1: strongly disagree to 5: strongly agree and an additional "I don't know" option. We instantiated the selected questions according to the purpose and content of the Q-Rapids quality model. At the time of the evaluation, the Q-Rapids quality model provided product owners with support for identifying product and process deficiencies.

*B. Execution*

In December 2017, we installed the Q-Rapids tool in each company. We gathered product and process data in each company for at least two weeks. Then we evaluated the Q-Rapids quality model in January 2018 following the procedures described above.

*C. Data Analysis*

We first carried out within-case analyses of the quantitative and qualitative data for each company. Then we compared, correlated, and integrated the results among the companies (cross-case analyses) [22].

We report descriptive statistics including the sample size (N), median (Mdn), minimum (Min), maximum (Max), and modal value (Mode) for the quantitative analyses. Regarding the qualitative analysis, we used data-driven thematic analysis [23] to analyze the participants' feedback on the Q-Rapids quality model. We inductively derived themes (i.e., we explicitly mentioned suggestions for improvement) by coding and interpreting all observation protocols.

*D. Results*

In total, three product owners, four project managers, and one developer across the four companies participated in the evaluation. They all had at least three years of work experience in their companies (Mdn = 10.5, Min = 3, Max = 30) and at least half a year of work experience in their current role (Mdn = 7, Min = 0.5, Max = 30).

The majority of the participants claimed that the assessed metrics included in the implemented Q-Rapids quality model are understandable (N = 7, Mdn = 4, Min = 2, Max = 5, Mode = 4). One participant pointed out: *"I need more clarification and details"* to understand the assessed metrics. S/he and other participants suggested adding the actual values of the metrics (i.e., raw values before normalization via utility functions). In general, most of the participants had difficulties understanding the normalized values, as stated by one participant: *"I don't understand […] what it means. I know 1 is good and 0 is bad, but what about when it is 0.91?"* Moreover, they perceived the understandability of the product and process factors as moderately understandable (N = 7, Mdn = 3, Min = 2, Max = 5, Mode = 2.5). In general, the participants proposed avoiding negated formulations of the factors or metrics in order to increase their understandability; for instance, using *complex files* instead of *non-complex files*.

All participants agreed that the current Q-Rapids quality model is relevant for their work (N = 7, Mdn = 4, Min = 3, Max = 4.5, Mode = 4). They recommended linking the provided information about the product and process factors as well as the assessed metrics with further information sources (e.g., issue reports) in order to better support the decision-making process. The participants agreed that integrating several heterogeneous data sources is an added value for supporting actionable analytics in their companies.

*E. Implications of the Results: Ongoing Work*

One of the most important suggestions for improvement was to visualize raw data to facilitate decision-making. As explained in Fig. 1, the three most abstract levels (i.e., quality aspects, product factors, and assessed metrics) work as "traffic lights", with a normalized value between 0 and 1 and customized thresholds. Bearing in mind the evaluation results, we plan to train stakeholders to explore the raw data in Kibana so they can look for the problem and make a decision when a quality alert is raised. Below, we provide another example on how to use our quality model with respect to the *reliability* quality aspect including actionable analytics:

At the end of the last cycle, a new release of the product was launched. Unfortunately, during the current cycle the traffic light for *reliability* suddenly turned red and an alarm was triggered. The Q-Rapids quality model showed Bob orange for *testing status* and red for *software stability*. He called an immediate meeting with Jane. Together, they recognized that the problem was related to *errors at runtime*. Jane further explored the raw data from the logs and noticed many critical errors (e.g., 5xx errors) caused by one module. Therefore, immediate action was required to solve the crashes and exceptions identified at the client side. Using further drill-down, Jane was able to identify the exact line of code responsible for the mess and informed her team to work on the issue. Going further, Jane realized that the *test coverage* of this module was worse than average. Therefore, the product owner was able to identify it and include it in the product backlog with lower priority. The *errors at runtime* metric was recovered within one hour, whereas the *test coverage* got a stable desired value a few days later.

*F. Threats to validity*

We developed and evaluated the Q-Rapids quality model drawing on a convenient sample of product owners and managers (*Sample Bias*). Thus, our results are tied to the context of the elicited use cases and the companies involved in the Q-Rapids project. To mitigate the risk of *social desirability*, we informed all participants that this evaluation was being performed at an early stage of the Q-Rapids project to get early feedback on the Q-Rapids quality model and support the development of the Q-Rapids tool. Moreover, the evaluation included only one treatment – the Q-Rapids quality model – (*Mono-Operation Bias*). Therefore, the results can only be interpreted as an indication of the understandability and relevance of the Q-Rapids quality model. The Q-Rapids quality model might serve as a basis for supporting actionable analytics in companies developing their RSD projects in a similar setting. Further evaluations in different company settings, including a larger sample of decision makers and alternative treatments, are required in order to generalize the results to other organizations applying RSD.

## VII. CONCLUSIONS AND FUTURE WORK

There is a need for making software development decisions and their rationale available to all project members. If tacit knowledge is replaced by a tangible quality model, it is possible to raise quality alerts on a real-time basis based on the measurement and analysis of software quality, development productivity, and software usage. These quality alerts enable actionable analytics. These actions can be added to the product backlog of the next RSD cycle and, hence, increase the transparency of decision-making [10].

In this paper, we presented three main contributions: (i) a quality model combining heterogeneous data sources, elicited systematically together with practitioners (c.f. Table I); (ii) the implementation of the quality model based on Big Data technologies, including frameworks for collecting and analyzing data. Other organizations can use the implemented quality model to gather and analyze information automatically and manage software quality in each RSD cycle. The proposed solution will be released in 2019 on our website: http://q-rapids.eu/. (iii) The preliminary evaluation of the quality model indicates that the assessed metrics and the implemented factors are understandable. Yet, their understandability can be further improved, e.g., by additionally providing the actual values without normalizations. All participants assessed the quality model as relevant for their work and as supporting actionable analytics within their companies.

Despite the diverse contexts across the four companies we worked with, we observed an overlap regarding the available data and quality issues to be addressed when managing software quality. We conclude that it is possible to create a tailorable quality model for managing software quality in different RSD settings. Such a quality model has to be adapted to the company and to the project-specific context. Moreover, some interviewees in our study believe that the quality model introduced in Table I could be used as a benchmark for comparing the quality of competing software systems in a shared domain.

Future work will target several directions, mainly related to learning data in order to add new elements to the quality model, improving utility functions, identifying correlations, further customizing weights, as well as exploiting parallelism with Big Data analysis technologies.

## ACKNOWLEDGMENTS

This work has received support from the European Union's Horizon 2020 program under grant n° 732253. We thank all members of Bittium, ITTI, Nokia, and Softeam who participated in the workshops for creating and evaluating the Q-Rapids quality model as well as the company representatives (Sanja Aaramaa, Rafał Kozik, Jari Partanen, and Andrey Sadovykh) for supporting our research. We also thank Axel Wickenkamp for implementing the Q-Rapids quality model as well as Lidia López, Woubshet Behutiye, and Pertti Karhapää for supporting the evaluation execution.